\documentclass[11pt]{article}
\usepackage{amsmath,amssymb}

\newcommand{\ba}{\begin{alignat}{3}}
\newcommand{\e}{\epsilon}

\newcommand{\dl}{\delta}

\newcommand{\g}{\gamma}

\newcommand{\sg}{\sigma}

\newcommand{\pa}{\partial}

\newcommand{\om}{\omega}
\newcommand{\mc}{\mathcal}

\begin{document}

\begin{titlepage}
\begin{flushright}
\end{flushright}
\begin{center}
  \vspace{4cm}
  {\bf \Large Super Virasoro Algebra From Supergravity}
  \\  \vspace{2cm}
  Yoshifumi Hyakutake
   \\ \vspace{1cm}
   {\it College of Science, Ibaraki University \\
   Bunkyo 1-1, Mito, Ibaraki 310-0062, Japan}
\end{center}

\vspace{2cm}
\begin{abstract}
We investigate AdS$_3$/CFT$_2$ correspondence in three dimensional supergravity.
We construct a current for general coordinate invariance and 
that for local supersymmetry via covariant approach. 
Hamiltonian and supercharge are well defined in terms of vielbein and spin connection.
After discussing asymptotic supersymmetry group at the boundary of AdS$_3$ geometry,
we show that there exist a direct product of Virasoro algebras at the boundary.
We also show that one of them is extended to super Virasoro algebra.
\end{abstract}
\end{titlepage}

\setlength{\baselineskip}{0.65cm}

\section{Introduction}

One of important directions in string theory is the understanding of the correspondence
between gauge theory and gravity theory\cite{malda}.
From this correspondence, it is possible to predict physical quantities from the gravity side, 
such as correlation functions in the gauge theory at strong coupling limit\cite{witten1,gkp}. 
Although the gauge/gravity correspondence is not proved yet, it is widely applied to various fields and 
we get some insight into the gauge theory at strong coupling region.

Among many works which support the gauge/gravity correspondence,
it is very important to examine AdS$_3$/CFT$_2$ correspondence since both three dimensional gravity
with negative cosmological constant and two dimensional conformal field theory are deeply investigated.
In three dimensional gravity theory with negative cosmological constant, a vacuum solution
which has negative energy is described by global AdS$_3$ geometry\cite{deser}.
And there are so-called BTZ black hole solutions as excited configurations\cite{btz,bhtz}.
In two dimensional CFT, a number of generators of the symmetry becomes infinite and they form a
Virasoro algebra\cite{bpz}.
In 1986, Brown and Henneaux have showed that there exist Virasoro
algebras at the boundary of AdS$_3$ geometry without employing string theory\cite{brown}. 
Applying this result, Strominger has succeeded to explain an entropy of BTZ black hole
via Cardy formula\cite{stro}.
Another important approach to the three dimensional gravity is the relationship 
between the gravity theory and the gauge Chern-Simons theory\cite{achu,witten2}.
And it is possible to derive the Virasoro algebras at the boundary of the gauge 
Chern-Simons theory\cite{bana,carl}.

The purpose of this paper is investigate AdS$_3$/CFT$_2$ correspondence in three dimensional supergravity.
It is expected that the Virasoro algebra at the boundary of AdS$_3$ will be extended to super Virasoro algebra.
In fact, Banados et al. have showed the existence of the super Virasoro algebra
by using supersymmetric gauge Chern-Simons theory\cite{bana2}.
In this paper, we construct a current for the general covariance and that for the local supersymmetry
in terms of vielbein and spin connection in a covariant way.
In order to execute this, we employ Noether's method and Wald's covariant approach\cite{crn,lee,wald1}.

We also examine asymptotic supersymmetry group which preserves the boundary
behaviors of the vielbein and the spin connection.
We confirm that global AdS$_3$ geometry corresponds to the ground state of Neveu-Schwarz sector and 
massless BTZ black hole does to that of Ramond sector\cite{cou}.

Finally we evaluate variations of Hamiltonian and supercharge.
These variations should close up to central extensions. 
We will explicitly calculate the central extensions and show that there exists
a direct product of super Virasoro algebra and Virasoro algebra at the boundary.
Central charge correctly matches with the results obtained in previous works.

Organization of the paper is as follows.
In section 2, we check the general coordinate invariance and the local supersymmetry
in three dimensional $\mathcal{N}=(1,0)$ supergravity.
In section 3, the Noether current for the general coordinate invariance is constructed,
and in section 4, that for the local supersymmetry is defined.
We discuss properties of the asymptotic supersymmetry group by solving relaxed
Killing vector and Killing spinor equations.
Finally in section 6 we discuss Virasoro algebras realized at the boundary.
Section 7 is devoted to conclusion and discussion. There are three appendices
to support some technical calculations.

\section{Supergravity Lagrangian and Local Symmetry}

In this section we consider three dimensional supergravity which consists of 
a vielbein $e^a{}_\mu$ and a Majorana gravitino $\psi_\mu$. 
$\mu, \nu$ are used for space-time indices and $a,b$ are for local Lorentz ones.
Both fields contain no physical degrees of freedom, so it is possible to construct
the action which possesses local supersymmetry.

For simplicity, we consider $\mathcal{N}=(1,0)$ supergravity.
Since our main interest is AdS$_3$/CFT$_2$ correspondence, we also introduce
negative cosmological constant $\Lambda=-\frac{1}{\ell^2}$ and its super partner. 
Then the action $S$ or the Lagrangian $\mathcal{L}$ of three dimensional supergravity 
with negative cosmological constant is given by
\ba
  S = \int d^3 x \, \mathcal{L}
  \sim \frac{1}{16\pi G_\text{N}} \int d^3x \, e \Big( R + \frac{2}{\ell^2} 
  - \frac{1}{2} \overline{\psi_\rho} \g^{\mu\nu\rho} \psi_{\mu\nu} \Big), \label{eq:SGLag}
\end{alignat}
where $G_\text{N}$ is the gravitational constant in three dimensions.
A matrix $\gamma^{\rho\mu\nu}$ is a completely antisymmetric tensor constructed 
by gamma matrix $\gamma^\mu$.
Notations and some useful relations of the gamma matrix in three dimensions are 
summarized in the appendix \ref{app:gamma}.
The symbol $\sim$ will be used when higher order terms on $\psi_\mu$ are neglected.
In the above equation, $\mathcal{O}(\psi^4)$ terms are ignored.
Through this paper, we use $D_\mu$ for the covariant derivative which acts 
only on local Lorentz indices. For example, 
$D_\mu \psi_\nu = \partial_\mu \psi_\nu + \tfrac{1}{4} \om_{\mu ab} \g^{ab} \psi_\nu$,
where $\omega_{\mu ab}$ is a spin connection.
A curly covariant derivative $\mathcal{D}_\mu$ is defined like
\ba
  \mathcal{D}_\mu \psi_\nu &= D_\mu \psi_\nu + \frac{1}{2\ell} \gamma_\mu \psi_\nu. \label{eq:covD}
\end{alignat}
A field strength $\psi_{\mu\nu}$ of the Majorana gravitino is defined by using 
the curly covariant derivative as $\psi_{\mu\nu} \equiv 2 \mathcal{D}_{[\mu} \psi_{\nu]}$.

The action~(\ref{eq:SGLag}) is invariant under the general coordinate transformation and 
local supersymmetry. First, let us check the general coordinate invariance.
Under general coordinate transformation $x'^\mu = x^\mu - \xi^\mu$, 
the vielbein and the Majorana gravitino transform like vector fields,
\ba
  \dl_\xi e^a{}_\mu = \xi^\nu \pa_\nu e^a{}_\mu + \pa_\mu \xi^\nu e^a{}_\nu, \qquad
  \dl_\xi \psi_\mu = \xi^\nu \pa_\nu \psi_\mu + \pa_\mu \xi^\nu \psi_\nu. \label{eq:GRtr}
\end{alignat}
Then $\dl_\xi e = e e^\mu{}_a \dl_\xi e^a{}_\mu = \pa_\mu(e\xi^\mu)$,
and the Lagrangian density $e^{-1} \mathcal{L}$ behaves as a scalar field.
Therefore the variation of the Lagrangian under the general coordinate transformation becomes
\ba
  \dl_\xi \mathcal{L} = \pa_\mu \big( \xi^\mu \mathcal{L} \big). \label{eq:SGtr}
\end{alignat}
Thus the action (\ref{eq:SGLag}) is invariant under the general coordinate transformation.

Next let us check $\mathcal{N}=(1,0)$ local supersymmetry of the action~(\ref{eq:SGLag})
up to $\mathcal{O}(\psi^2)$. 
Under the local supersymmetry, the vielbein and the Majorana gravitino transform like
\ba
  \dl_\e e^a{}_\mu &= \overline{\e} \g^a \psi_\mu, \qquad
  \dl_\epsilon \psi_\mu &= 2 \mathcal{D}_\mu \e. \label{eq:susyvariation}
\end{alignat}
Here $\epsilon(x)$ represents a space-time dependent parameter which belongs to 
a Majorana representation. Then the variation of the field strength of the Majorana gravitino
is evaluated as
\ba
  \dl_\epsilon \psi_{\mu\nu} &\sim 2 [\mathcal{D}_\mu, \mathcal{D}_\nu] \e \notag
  \\
  &= \frac{1}{2} R_{ab\mu\nu} \gamma^{ab} \e + \frac{1}{\ell^2} \gamma_{\mu\nu} \e. \label{eq:susytrp2}
\end{alignat}
In the first line, the variations of the vielbein and the spin connection are neglected
since these give $\mathcal{O}(\psi^2)$ terms. 
By using eq.~(\ref{eq:susytrp2}) and employing relations of gamma matrices given 
in the appendix \ref{app:gamma}, the variation of the fermionic bilinear term in (\ref{eq:SGLag}) becomes
\ba
  &\dl_\epsilon \Big( - \frac{1}{2} e \overline{\psi_\rho} \g^{\mu\nu\rho} \psi_{\mu\nu} \Big) \notag
  \\
  &\sim - \partial_\rho \big( e \overline{\epsilon} \g^{\mu\nu\rho} \psi_{\mu\nu} \big)
  + e \overline{\epsilon} \g^{\mu\nu\rho} \mathcal{D}_\rho \psi_{\mu\nu} 
  - e \overline{\psi_\rho} \g^{\mu\nu\rho} 
  \Big(\frac{1}{4} R_{ab\mu\nu} \gamma^{ab} + \frac{1}{2\ell^2} \gamma_{\mu\nu} \Big) \e \notag
  \\
  &= - \partial_\rho \big( e \overline{\epsilon} \g^{\mu\nu\rho} \psi_{\mu\nu} \big)
  + \frac{1}{4} e R_{ab\mu\nu} \overline{\epsilon} \{ \g^{\mu\nu\rho}, \gamma^{ab} \} \psi_{\rho} 
  + \frac{1}{2\ell^2} e \overline{\epsilon} \{ \g^{\mu\nu\rho}, \gamma_{\mu\nu} \} \psi_{\rho} \notag
  \\
  &= \partial_\mu \big( e \overline{\epsilon} \Psi^\mu \big)
  + 2 e G^a{}_\mu \overline{\epsilon} \g^\mu \psi_a, \label{eq:varf}
\end{alignat}
where
\begin{alignat}{3}
  G {}^a{}_\mu &= R^a{}_\mu - \frac{1}{2} e^a{}_\mu \Big( R + \frac{2}{\ell^2} \Big), \notag
  \\
  \Psi {}^\rho &= - \g^{\mu\nu\rho} \psi_{\mu\nu} = - \e^{\mu\nu\rho} \psi_{\mu\nu},
\end{alignat}
are equations of motion for $e^\mu{}_a$ and $\psi_\rho$, respectively.
Since the variation of the fermionic bilinear term is obtained up to $\mathcal{O}(\psi^2)$,
it will be cancelled by that of purely bosonic terms in the Lagrangian (\ref{eq:SGLag}).
The supersymmeric transformation of the purely bosonic terms in (\ref{eq:SGLag}) is evaluated like
\ba
  \dl_\e \Big\{ e \Big( R + \frac{2}{\ell^2} \Big) \Big\}
  &\sim - 2 e G^a{}_\mu \overline{\epsilon} \g^\mu \psi_a
  + 2 e e^\mu{}_a e^\nu{}_b D_\mu \dl_\e \omega_\nu{}^{ab} \notag
  \\
  &= - 2 e G^a{}_\mu \overline{\epsilon} \g^\mu \psi_a
  + \pa_\mu ( 2 e e^\mu{}_a e^\nu{}_b \dl_\e \omega_\nu{}^{ab})
  - D_\mu (2 e e^\mu{}_a e^\nu{}_b) \dl_\e \omega_\nu{}^{ab}. \label{eq:varb}
\end{alignat}
Notice that the third term in the last line should be used to express
the spin connection in terms of the vielbein and the Majorana gravitino.
Details can be found in appendix \ref{app:spin}. 
So the third term is neglected here and combining eq.~(\ref{eq:varf}) and eq.~(\ref{eq:varb}),
the variation of the Lagrangian under the local supersymmetry becomes
\ba
  16 \pi G_\text{N} \dl_\e \mathcal{L} \sim
  \pa_\mu \big( 2 e e^\mu{}_a e^\nu{}_b \dl_\e \om_\nu{}^{ab} + e \overline{\epsilon} \Psi^\mu \big).
  \label{eq:SGvarsusy}
\end{alignat}
Thus the action (\ref{eq:SGLag}) is invariant under the local supersymmetry.

\section{Current for the General Covariance}
\label{sec:currentxi}

In this section, we construct a current for the general coordinate transformation, $x'^\mu = x^\mu - \xi^\mu$. 
In order to do this, we employ a covariant approach which is investigated in refs.~\cite{crn,lee,wald1}.
In stead of the metric, we regard the vielbein as a fundamental field since 
the Lagrangian (\ref{eq:SGLag}) is written in terms of the vielbein and the spin connection.
The current is constructed up to $\mathcal{O}(\psi^3)$.

First let us consider the variation of the Lagrangian (\ref{eq:SGLag}).
As in the previous section, we neglect the equation of motion for the spin connection
which is solved to express the spin connection in terms of the vielbein and the Majorana gravitino.
(See appendix~\ref{app:spin}.)
Then the variation of the Lagrangian becomes
\ba
  16 \pi G_\text{N} \dl \mathcal{L}
  = 2e G^a{}_\mu \dl e^\mu{}_a + e \overline{\dl \psi_\mu} \Psi^\mu 
  + \pa_\mu \big(e \Theta^\mu(\dl) \big), \label{eq:SGvariation}
\end{alignat}
where we defined
\begin{alignat}{3}
  \Theta^\mu(\dl) = 2 e^\mu{}_a e^\nu{}_b \dl \om_\nu{}^{ab} 
  + \overline{\psi_\nu} \gamma^{\mu\nu\rho} \dl \psi_\rho. \label{eq:theta}
\end{alignat}
Now we identify the variation with that of general coordinate transformation~(\ref{eq:GRtr}).
By using the relations
\begin{alignat}{3}
  \dl_\xi \om_\nu{}^{ab} &= \xi^\rho \pa_\rho \om_\nu{}^{ab} + \pa_\nu \xi^\rho \om_\rho{}^{ab} \notag
  \\
  &= \xi^\rho R^{ab}{}_{\rho\nu} + D_\nu (\xi^\rho \om_\rho{}^{ab}),
  \\
  \dl_\xi \psi_\rho &= \xi^\sg \pa_\sg \psi_\rho + \pa_\rho \xi^\sg \psi_\sg \notag
  \\
  &= \xi^\sg \psi_{\sg\rho} + \mathcal{D}_\rho (\xi^\sg \psi_\sg)
  - \xi^\sg \Big( \frac{1}{4} \om_{\sg ab} \g^{ab} + \frac{1}{2\ell} \g_\sg \Big) \psi_\rho,
\end{alignat}
$\Theta^\mu(\xi) \equiv \Theta^\mu(\dl_\xi)$ is written as
\begin{alignat}{3}
  e \Theta^\mu(\xi) &= 2 e R^\mu{}_\nu \xi^\nu 
  + 2 e e^\mu{}_a e^\nu{}_b D_\nu (\xi^\rho \om_\rho{}^{ab}) \notag
  \\
  &\quad\,
  + e \xi^\sg \overline{\psi_\nu} \gamma^{\mu\nu\rho} \psi_{\sg\rho} 
  + e \overline{\psi_\nu} \gamma^{\mu\nu\rho} \mathcal{D}_\rho (\xi^\sg \psi_\sg)
  - e \xi^\sg \overline{\psi_\nu} \gamma^{\mu\nu\rho} 
  \Big( \frac{1}{4} \om_{\sg ab} \g^{ab} + \frac{1}{2\ell} \g_\sg \Big) \psi_\rho \notag
  \\
  &= 2 e R^\mu{}_\nu \xi^\nu 
  - \Big\{ D_\nu \big( 2 e e^\mu{}_a e^\nu{}_b \big) 
  + \frac{1}{4} e \overline{\psi_\nu} \gamma^{\mu\nu\sigma} \g_{ab} \psi_\sigma \Big\} 
  \xi^\rho \om_\rho{}^{ab} \notag
  \\
  &\quad\,
  + \pa_\nu \big( 2e e^\mu{}_a e^\nu{}_b \xi^\rho \om_\rho{}^{ab} 
  + e \xi^\sigma \overline{\psi_\rho} \g^{\mu\rho\nu} \psi_\sigma \big) \notag
  \\
  &\quad\,
  + e \xi^\sg \overline{\psi_\nu} \gamma^{\mu\nu\rho} \psi_{\sg\rho}
  - \frac{1}{2\ell} e \xi^\sg \overline{\psi_\nu} \gamma^{\mu\nu\rho} \g_\sg \psi_\rho 
  - e \xi^\sg \overline{\mathcal{D}_\rho \psi_\nu} \gamma^{\mu\nu\rho} \psi_\sg \notag
  \\
  &= \pa_\nu \big( 2e e^\mu{}_a e^\nu{}_b \xi^\rho \om_\rho{}^{ab} 
  + e \xi^\sigma \overline{\psi_\rho} \g^{\mu\rho\nu} \psi_\sigma \big) \notag
  \\
  &\quad\,
  + 2 e \Big( R^\mu{}_\nu - \frac{1}{4\ell} \overline{\psi_\rho} 
  \g^{\mu\rho\sigma}\g_\nu \psi_\sigma \Big) \xi^\nu
  + e \xi^\sg \Big( \overline{\psi_\nu} \g^{\mu\nu\rho} \psi_{\sg\rho}
  + \frac{1}{2} \overline{\psi_\sg} \g^{\mu\nu\rho} \psi_{\nu\rho} \Big) \notag
  \\
  &= \pa_\nu \big( 2e e^\mu{}_a e^\nu{}_b \xi^\rho \om_\rho{}^{ab} 
  + e \xi^\sigma \overline{\psi_\rho} \g^{\mu\rho\nu} \psi_\sigma \big) 
  + 16\pi G_\text{N} \xi^\mu \mathcal{L} \notag
  \\
  &\quad\,
  + 2 e \tilde{G}^\mu{}_\nu \xi^\nu
  + e \xi^\sg \Big( \overline{\psi_\nu} \g^{\mu\nu\rho} \psi_{\sg\rho}
  + \frac{1}{2} \overline{\psi_\sg} \g^{\mu\nu\rho} \psi_{\nu\rho} \Big). \label{eq:thetaxi}
\end{alignat}
The second term in the third line vanishes after expressing the spin connection
in terms of the vielbein and the Majorana gravitino.
Since $\psi_{\mu\nu} = \frac{1}{2} \e_{\mu\nu\rho} \Psi^\rho$, the last line also
vanishes after imposing the equations of motion, $\tilde{G}^\mu{}_\nu = 0$ and $\psi_{\mu\nu}=0$.
Here $\tilde{G}^\mu{}_\nu$ includes fermionic bilinear term and its explicit form is given
in the appendix~\ref{app:spin}.

Let us apply Noether's procedure to construct a current for general coordinate invariance.
Subtracting eq.~(\ref{eq:SGtr}) from eq.~(\ref{eq:SGvariation}), it is possible to define
the current up to $\mathcal{O}(\psi^3)$ as
\ba
  &16\pi G_\text{N} e J^\mu(\xi) \notag
  \\
  &= e \Theta^\mu (\xi) - 16\pi G_\text{N} \xi^\mu \mathcal{L} 
  + \pa_\nu \big( e \tilde{Q}^{\mu\nu}(\xi) \big) \label{eq:SGcurrent}
  \\
  &= \pa_\nu \big\{ e \big( Q^{\mu\nu}(\xi) + \tilde{Q}^{\mu\nu}(\xi) \big) \big\}
  + 2 e \tilde{G}^\mu{}_\nu \xi^\nu
  + e \xi^\sg \Big( \overline{\psi_\nu} \g^{\mu\nu\rho} \psi_{\sg\rho}
  + \frac{1}{2} \overline{\psi_\sg} \g^{\mu\nu\rho} \psi_{\nu\rho} \Big), \notag
\end{alignat}
where 
\begin{alignat}{3}
  Q^{\mu\nu}(\xi) = 2 e^\mu{}_a e^\nu{}_b \xi^\rho \om_\rho{}^{ab} 
  + \xi^\sigma \overline{\psi_\rho} \g^{\mu\rho\nu} \psi_\sigma.
\end{alignat}
$\tilde{Q}^{\mu\nu}(\xi)$ is some antisymmetric tensor
which is necessary to make the variation of the Hamiltonian well defined.
In order to fix the form of $\tilde{Q}^{\mu\nu}(\xi)$, let us examine the variation 
of the current (\ref{eq:SGcurrent}).
\ba
  &\dl \big( 16\pi G_\text{N} e J^\mu(\xi) \big) \notag
  \\
  &= \dl \big( e \Theta^\mu(\xi) \big) - 16\pi G_\text{N} \xi^\mu \dl \mathcal{L}
  + \pa_\nu \big\{ \dl \big( e \tilde{Q}^{\mu\nu}(\xi) \big) \big\} \notag
  \\
  &= \dl \big( e \Theta^\mu(\xi) \big) 
  - \xi^\mu \partial_\nu \big( e \Theta^\nu(\dl) \big) 
  - 2 e \xi^\mu G^a{}_\nu \dl e^\nu{}_a - e \xi^\mu \overline{\dl \psi_\nu} \Psi^\nu
  + \pa_\nu \big\{ \dl \big( e \tilde{Q}^{\mu\nu}(\xi) \big) \big\} \notag
  \\
  &= e \om^\mu(\xi,\dl) + \partial_\nu \big\{ e \big( \xi^\nu \Theta^\mu(\dl) 
  - \xi^\mu \Theta^\nu(\dl) \big) + \dl \big( e \tilde{Q}^{\mu\nu}(\xi) \big) \big\}
  - 2 e \xi^\mu G^a{}_\nu \dl e^\nu{}_a - e \xi^\mu \overline{\dl \psi_\nu} \Psi^\nu,  \label{eq:varSG}
\end{alignat}
where
\ba
  e \om^\mu(\xi,\dl) = \dl \big( e \Theta^\mu(\xi) \big) 
  - \dl^\text{L}_{\xi} \big( e \Theta^\mu(\dl g) \big).
\end{alignat}
Notice that $\dl^\text{L}_\xi$ represents the Lie derivative along $\xi$ direction,
and the relation $\dl^\text{L}_{\xi} \big( e \Theta^\mu(\dl) \big) = 
\pa_\nu \big( e \xi^\nu \Theta^\mu(\dl) \big) - e \pa_\nu \xi^\mu \Theta^\nu(\dl)$
is used to derive the last equation in eq.~(\ref{eq:varSG}).
$\om^\mu(\dl,\xi)$ is called the symplectic current and antisymmetric under the exchange of $\dl$ and $\dl_\xi$.
The integral of its time component corresponds to the variation of the Hamiltonian.
In order to make this variation well defined, we require the cancellation of the total divergent term like
\ba
  \dl \big( e \tilde{Q}^{\mu\nu}(\xi) \big) = 
  e \big( \xi^\mu \Theta^\nu(\dl) - \xi^\nu \Theta^\mu(\dl) \big).
\end{alignat}
It is not obvious whether we can find $\tilde{Q}^{\mu\nu}(\xi)$ which satisfies
the above relation. However, the existence of $\tilde{Q}^{\mu\nu}(\xi)$ is confirmed
by examining an integrability condition in ref. \cite{koga}.
As a summary, the variation of the current for the general coordinate invariance is given by
\ba
  \dl \big( 16\pi G_\text{N} e J^\mu(\xi) \big) 
  &= \pa_\nu \big\{ \dl \big(e Q^{\mu\nu}(\xi) \big) 
  + e \big( \xi^\mu \Theta^\nu(\dl) - \xi^\nu \Theta^\mu(\dl) \big) \big\} \label{eq:varSGcurrent}
  \\
  &\quad\,
  + \dl \Big \{ 2 e \tilde{G}^\mu{}_\nu \xi^\nu
  + e \xi^\sg \Big( \overline{\psi_\nu} \g^{\mu\nu\rho} \psi_{\sg\rho}
  + \frac{1}{2} \overline{\psi_\sg} \g^{\mu\nu\rho} \psi_{\nu\rho} \Big) \Big\}. \notag
\end{alignat}
The last line always contain equations of motion or their variations.
Therefore, as far as we consider variations along the moduli space of classical solutions,
the last line can be neglected.

\section{Current for the Local Supersymmetry}
\label{sec:currente}

In this section, we construct a Noether current for the local supersymmetry up to $\mathcal{O}(\psi^2)$.
The variation of the Lagrangian (\ref{eq:SGLag}) with equations of motion is 
given in eq.~(\ref{eq:SGvariation}), so we identify the variation with the 
supersymmetric transformation $\dl_\e$ of eq.~(\ref{eq:susyvariation}).
Then the total derivative term $\Theta(\e) \equiv \Theta(\dl_\e)$ becomes
\ba
  e \Theta^\mu(\e) &= 2 e e^\mu{}_a e^\nu{}_b \delta_\e \omega_{\nu}{}^{ab} 
  + 2 e \overline{\psi_\nu} \g^{\mu\nu\rho} \mathcal{D}_\rho \e \notag
  \\
  &= 2 e e^\mu{}_a e^\nu{}_b \delta_\e \omega_{\nu}{}^{ab} 
  + \pa_\rho \big( 2 e \overline{\psi_\nu} \g^{\mu\nu\rho} \e \big)
  - 2 e \overline{\mathcal{D}_\rho \psi_\nu} \g^{\mu\nu\rho} \e \notag
  \\
  &= 2 e e^\mu{}_a e^\nu{}_b \delta_\e \omega_{\nu}{}^{ab} 
  - \pa_\nu \big( 2 e \overline{\e} \g^{\mu\nu\rho} \psi_\rho \big)
  - e \overline{\e} \Psi^\mu. \label{eq:thetae}
\end{alignat}
The variation of the Lagrangian (\ref{eq:SGLag}) under the local supersymmetry (\ref{eq:susyvariation}) 
is already evaluated as in eq.~(\ref{eq:SGvarsusy}).
Subtracting eq.~(\ref{eq:SGvarsusy}) from eq.~(\ref{eq:thetae}), we can construct
the current for the local supersymmetry as
\begin{alignat}{3}
  16\pi G_\text{N} e S {}^\mu(\e) 
  &= e \Theta^\mu(\e) 
  - \big( 2 e e^\mu{}_a e^\nu{}_b \dl_\e \om_\nu{}^{ab} + e \overline{\epsilon} \Psi^\mu \big) \notag
  \\
  &= \partial_\nu \big( e U^{\mu\nu}(\e) \big)
  - 2 e \overline{\e} \Psi {}^\mu, \label{eq:SGsupercurrent}
\end{alignat}
where
\begin{alignat}{3}
  U^{\mu\nu}(\e) &= - 2 \overline{\e} \g^{\mu\nu\rho} \psi_\rho 
  = - 2 \e^{\mu\nu\rho} \overline{\e} \psi_\rho.
\end{alignat}
The super current is expressed as a total derivative term up to the equations of motion.
In order to examine whether the current is well defined or not, 
let us consider the supersymmetric transformation of the super current up to $\mathcal{O}(\psi^2)$.
\begin{alignat}{3}
  16\pi G_\text{N} \dl_{\e_2} \big( e S {}^\mu(\e_1) \big)
  &\sim \partial_\nu \big( - 4 e \e^{\mu\nu\rho} \overline{\e_1} \mathcal{D}_\rho \e_2 \big) 
  + e \overline{\e_1} \g^{\mu\nu\rho} \Big( R_{ab\nu\rho} \g^{ab} + \frac{2}{\ell^2} \g_{\nu\rho} \Big) 
  \e_2 \notag
  \\
  &= \partial_\nu \big\{ 2 e \e^{\mu\nu\rho} \big( \overline{\e_2} \mathcal{D}_\rho \e_1
  - \overline{\e_1} \mathcal{D}_\rho \e_2 \big) \big\} 
  + 4 e G^\mu{}_\nu \overline{\e_1} \g^{\nu} \e_2. \label{eq:varSGsuper}
\end{alignat}
The last equation is antisymmetric under the exchange of $\e_1$ and $\e_2$,
so it is consistent with the anticommutation relation of the supercharge.

\section{Asymptotic Supersymmetry Group for AdS$_3$ Geometry}
\label{sec:Killing}

In previous sections, we constructed currents for the general coordinate invariance 
and the local supersymmetry. We also derived the variations of the currents. 
Our next task is to evaluate these quantities at the boundary of AdS$_3$ geometry.
In order to do this, it is necessary to find a symmetry group which preserves the asymptotic
behavior of AdS$_3$ geometry, which is called asymptotic symmetry group. 
In our case, this group should be supersymmetric.

At spatial infinity, the metric of AdS$_3$ geometry becomes
\begin{alignat}{3}
  ds^2 = - N^2 dt^2 + r^2 d\phi^2 + N^{-2} dr^2, \quad N = \frac{r}{\ell}, \label{eq:AdS3}
\end{alignat}
where $r$ is a radial direction and $\phi$ is an angular one with $0 \le \phi \le 2\pi$.
This corresponds to the massless BTZ black hole.
Then the vielbein and the spin connection take the forms of
\begin{alignat}{3}
  &e^0 = N dt, \quad e^1 = r d\phi, \quad e^2 = N^{-1} dr, \notag
  \\
  &\omega^0{}_2 = N' e^0, \quad \omega^1{}_2 = \frac{N}{r} e^1,
\end{alignat}
where $a,b=0,1,2$ represent local Lorentz indices.
The isometry of this geometry becomes $SL(2,\mathbf{R}) \times SL(2,\mathbf{R})$.

Since we are only interested in the symmetry group at the boundary,
we investigate general coordinate transformation $x'^\mu = x^\mu - \xi^\mu$ 
which does not change the boundary behavior of AdS$_3$ geometry.
Then the condition to be imposed for the variation of the metric is as follows.
\ba
  \dl_\xi g_{\mu\nu} = 
  \begin{pmatrix}
    \mc{O}(1) & \mc{O}(1) & \mc{O}(r^{-1}) \\
    \mc{O}(1) & \mc{O}(1) & \mc{O}(r^{-1}) \\
    \mc{O}(r^{-1}) & \mc{O}(r^{-1}) & \mc{O}(r^{-4})
  \end{pmatrix},
\end{alignat}
where $\mu,\nu = t, \phi,r$.
The behaviors of diagonal components are determined so that these go to zero 
faster than eq.~(\ref{eq:AdS3}) as $r$ goes to infinity.
Then the behaviors of $\xi^\mu$ and off diagonal components around the boundary 
are simultaneously fixed.
The general coordinate transformation which satisfies the above condition is solved by
\ba
  \xi^t &= \ell \big( T_+(x^+) + T_-(x^-) \big), \notag
  \\
  \xi^\phi &= T_+(x^+) - T_-(x^-), \label{eq:xiasym}
  \\
  \xi^r &= - r \big( \pa_+ T_+(x^+) + \pa_- T_-(x^-) \big), \notag
\end{alignat}
where $x^\pm = \frac{t}{\ell} \pm \phi$ and $\pa_\pm = \frac{1}{2}(\ell \pa_t \pm \pa_\phi)$.
Thus the asymptotic symmetry group is parametrized by $T_+(x^+)$ or $T_-(x^-)$.

When $T_\pm(x^\pm)$ are substituted by Fourier modes of
\begin{alignat}{3}
  T_{\pm,n}(x^\pm)=\frac{1}{2}e^{inx^\pm}, 
\end{alignat}
the generators of asymptotic symmetry group are written as
\ba
  \xi_{\pm,n} = e^{inx^\pm} \pa_\pm - \frac{inr}{2} e^{inx^\pm} \pa_r. \label{eq:xiasym2}
\end{alignat}
These form direct sum of two Virasoro algebras,
$[\xi_{\pm,m}\,, \xi_{\pm,n}] = -i(m-n) \xi_{\pm,m+n}$.
Notice that $[\xi_{+,m}\,, \xi_{-,n}] = 0$ if there are no subleading terms in eq.~(\ref{eq:xiasym}).

Next let us examine a supersymmetric transformation $\epsilon$ which satisfy a boundary condition 
at spatial infinity. 
Because $\psi_\mu = 0$ for AdS$_3$ solution, the condition for the supersymmetric variation is imposed as
\ba
  \delta_\epsilon \psi_\mu = 
  \begin{pmatrix}
    \mc{O}(r^{-1/2}) & \mc{O}(r^{-1/2}) & \mc{O}(r^{-5/2}) 
  \end{pmatrix}. \label{eq:Killingspinor}
\end{alignat}
Now the gamma matrix is chosen as
\ba
  \g^0 = \begin{pmatrix} 0 & -1 \\ 1 & 0 \end{pmatrix}, \quad
  \g^1 = \begin{pmatrix} 0 & 1 \\ 1 & 0 \end{pmatrix}, \quad
  \g^2 = \begin{pmatrix} -1 & 0 \\ 0 & 1 \end{pmatrix}. \quad
\end{alignat}
Note that $\g^{012} = \mathbf{1}$ and $\g^{ab} = \e^{abc} \g_c$.
The Majorana fermion $\epsilon$ is decomposed into two Majorana-Weyl fermions in two dimensions,
$\epsilon_\pm = \frac{1}{2}(1\pm \g^2) \epsilon$, and these satisfy
$\g^2 \epsilon_{\pm} = \pm \epsilon_{\pm}$, $\g^1 \epsilon_{\pm} = \mp \g^0 \epsilon_{\pm}$
and $\g^{12} \epsilon_{\pm} = - \g^0 \epsilon_{\pm}$.
Then each component of the left hand side of eq.~(\ref{eq:Killingspinor}) is written as
\begin{alignat}{3}
  &\delta_\epsilon \psi_t
  = 2 \partial_t \epsilon - N N' \g^{02} \epsilon - \frac{N}{\ell} \g^0 \epsilon 
  = 2 \partial_t \epsilon_+ + 2 \partial_t \epsilon_- - \frac{2r}{\ell^2} \g^{0} \epsilon_+ , \notag
  \\
  &\delta_\epsilon \psi_\phi
  = 2 \partial_\phi \epsilon + N \g^{12} \epsilon + \frac{r}{\ell} \g^1 \epsilon 
  = 2 \partial_\phi \epsilon_+ + 2 \partial_\phi \epsilon_- - \frac{2r}{\ell} \g^{0} \epsilon_+ ,
  \\
  &\delta_\epsilon \psi_r
  = 2 \partial_r \epsilon + \frac{1}{\ell N} \g^2 \epsilon
  = 2 \partial_r \epsilon_+ + \frac{1}{r} \epsilon_+ 
  + 2 \partial_r \epsilon_- - \frac{1}{r} \epsilon_-. \notag
\end{alignat}
The solution of eq.~(\ref{eq:Killingspinor}) becomes
\begin{alignat}{3}
  \epsilon = r^{1/2} \g^0 \chi(x^+) + \ell r^{-1/2} \chi'(x^+), \label{eq:easym}
\end{alignat}
where $\chi(x^+)$ is a Majorana fermion with $\g^2 \chi = \chi$.
The solution depends only on $x^+$, so the remaining local supersymmetry is chiral in this sense.
Now we consider two solutions $\e_1$ and $\e_2$. 
Dirac conjugate of $\e_1$ is written as
$\overline{\e_1} =  - r^{1/2} \overline{\chi_1} \g^0 + \ell r^{-1/2} \overline{\chi'_1}$, 
so $\overline{\e_1}\e_2$ and $\overline{\e_1} \g^\mu \e_2$ are evaluated as
\ba
  \overline{\e_1} \e_2 &= 
  - \ell \, \overline{\chi_1} \g^0 \chi'_2 
  + \ell \, \overline{\chi'_1} \g^0 \chi_2, \notag
  \\
  \overline{\e_1} \g^t \e_2 &= \ell \, \overline{\chi_1} \g^0 \chi_2, \notag
  \\
  \overline{\e_1} \g^\phi \e_2 &= \overline{\chi_1} \g^0 \chi_2, 
  \\
  \overline{\e_1} \g^r \e_2 &= - r (\overline{\chi_1} \g^0 \chi_2)'. \notag
\end{alignat}
Notice that $\overline{\e_1} \g^\mu \e_2$ is proportional to the $x^+$ dependent part
of $\xi^{\mu}$, which is denoted as $\xi_+^\mu$, when 
$\overline{\chi_1} \g^0 \chi_2 = -i \chi_1^T \chi_2$ is proportional to $T_+$.

Let us consider the Fourier mode of $\e$, which is labelled as $\e_s$.
In order to fix the normalization of $\e_s$, we adopt the following relation
\begin{alignat}{3}
  \overline{\e_s} \g^\mu \e_t &= -2i \xi_{+,s+t}^\mu, \qquad
  \chi_s^T \chi_t = 2 T_{+,s+t}. \label{eq:normalization}
\end{alignat}
Then $\chi_s$ and $\e_s$ are fixed as
\ba
  \chi_s &= e^{i s x^+} \begin{pmatrix} 0 \\ 1 \end{pmatrix}, \qquad
  \epsilon_s = e^{i s x^+} 
  \begin{pmatrix}
    -r^{1/2} \\ i \ell s r^{-1/2} 
  \end{pmatrix}. \label{eq:easym2}
\end{alignat}
From eq.~(\ref{eq:normalization}), it is clear that $s+t$ should take some integer value.
When $s,t \in \mathbb{Z} + \frac{1}{2}$, those modes are in so called Neveu-Schwarz sector.
On the other hand, when $s,t \in \mathbb{Z}$, those modes are in Ramond sector.

Now let us calculate $\dl_\xi e^a{}_\mu = \xi^\rho \pa_\rho e^a{}_\mu + \pa_\mu \xi^\rho e^a{}_\rho$, 
because the currents are constructed in terms of the vielbein.
From eq.~(\ref{eq:xiasym}), $\dl_\xi e^a{}_\mu$ is evaluated as
\begin{alignat}{3}
  \dl_\xi e^a{}_\mu &= \begin{pmatrix}
    0 & r \big( \pa_+ T_+ - \pa_- T_- \big) & 0 \\
    \frac{r}{\ell} \big( \pa_+ T_+ - \pa_- T_- \big) & 0 & 0 \\
    - \pa_+^2 T_+ - \pa_-^2 T_- & - \ell \big( \pa_+^2 T_+ - \pa_-^2 T_- \big) & 0 
  \end{pmatrix}. \label{eq:dlxi}
\end{alignat}
It is, however, obvious that some off diagonal components do not go to zero faster that $e^a{}_\mu$
as $r$ goes to infinity. In order to avoid this problem, we employ local Lorentz transformation
$\dl_\Lambda e^a{}_\mu = \Lambda^a{}_b e^b{}_\mu$, where
\begin{alignat}{3}
  \Lambda^a{}_b &= \begin{pmatrix}
    0 & - \pa_+ T_+ + \pa_- T_- & \frac{\ell}{r} \big( \pa_+^2 T_+ + \pa_-^2 T_- \big) \\
    - \pa_+ T_+ + \pa_- T_- & 0 & - \frac{\ell}{r} \big( \pa_+^2 T_+ - \pa_-^2 T_- \big) \\
    \frac{\ell}{r} \big( \pa_+^2 T_+ + \pa_-^2 T_- \big) & \frac{\ell}{r} \big( \pa_+^2 T_+ - \pa_-^2 T_- \big) & 0
  \end{pmatrix}. \label{eq:Lambda}
\end{alignat}
By redefining $\dl_\xi e^a{}_\mu = \xi^\rho \pa_\rho e^a{}_\mu + \pa_\mu \xi^\rho e^a{}_\rho + \Lambda^a{}_b e^b{}_\mu$, 
we obtain
\begin{alignat}{3}
  \dl_\xi e^a{}_\mu &= \begin{pmatrix}
    0 & 0 & \frac{\ell^2}{r^2} \big( \pa_+^2 T_+ + \pa_-^2 T_- \big) \\
    0 & 0 & - \frac{\ell^2}{r^2} \big( \pa_+^2 T_+ - \pa_-^2 T_- \big) \\
    0 & 0 & 0
  \end{pmatrix}. \label{eq:dlxi2}
\end{alignat}
In a similar way, we define 
$\dl_\xi \om_\mu{}^{ab} = \xi^\rho \pa_\rho \om_\mu{}^{ab} + \pa_\mu \xi^\rho \om_\rho{}^{ab} 
+ \Lambda^a{}_c \om_\mu{}^{cb} + \Lambda^b{}_c \om_\mu{}^{ac}$.
After some calculations, the variation of the spin connection becomes
\begin{alignat}{3}
  \dl_\xi \om_t{}^a{}_b &= \begin{pmatrix}
    0 & 0 & - \frac{1}{r} \big( \pa_+^3 T_+ + \pa_-^3 T_- \big) \\
    0 & 0 & \frac{1}{r} \big( \pa_+^3 T_+ - \pa_-^3 T_- \big) \\
    - \frac{1}{r} \big( \pa_+^3 T_+ + \pa_-^3 T_- \big) & - \frac{1}{r} \big( \pa_+^3 T_+ - \pa_-^3 T_- \big) & 0
  \end{pmatrix}, \notag
  \\
  \dl_\xi \om_\phi{}^a{}_b &= \begin{pmatrix}
    0 & 0 & - \frac{\ell}{r} \big( \pa_+^3 T_+ - \pa_-^3 T_- \big) \\
    0 & 0 & \frac{\ell}{r} \big( \pa_+^3 T_+ + \pa_-^3 T_- \big) \\
    - \frac{\ell}{r} \big( \pa_+^3 T_+ - \pa_-^3 T_- \big) & - \frac{\ell}{r} \big( \pa_+^3 T_+ + \pa_-^3 T_- \big) & 0
  \end{pmatrix}, 
  \\
  \dl_\xi \om_r{}^a{}_b &= \begin{pmatrix}
    0 & 0 & \frac{\ell}{r^2} \big( \pa_+^2 T_+ + \pa_-^2 T_- \big) \\
    0 & 0 & - \frac{\ell}{r^2} \big( \pa_+^2 T_+ - \pa_-^2 T_- \big) \\
    \frac{\ell}{r^2} \big( \pa_+^2 T_+ + \pa_-^2 T_- \big) & \frac{\ell}{r^2} \big( \pa_+^2 T_+ - \pa_-^2 T_- \big) & 0
  \end{pmatrix}. \notag
\end{alignat}
The variation of the spin connection also goes to zero faster that $\om_\mu{}^a{}_b$.
These results will be employed to calculate central charges in the next section.

\section{Super Virasoro Algebra from Supergravity}
\label{sec:superVirasoro}

Now we are ready to construct super Virasoro algebra at the boundary of
the three dimensional supergravity.
As discussed in section \ref{sec:currentxi} and \ref{sec:currente}, we have constructed
the currents for the general coordinate invariance and the local supersymmetry.
Corresponding charges, the Hamiltonian and the supercharge, are obtained by integrating these currents
over two dimensional space.

First the Hamiltonian for the general coordinate transformation $\xi^\mu$ is given by
\ba
  H(\xi) &= \int d^2x \, e J^t(\xi) \notag
  \\
  &=\frac{1}{16\pi G_\text{N}} \oint_{r=\infty} \!\!\!\!\!\!\!\! d\phi \,
  \big( e Q^{tr}(\xi) + e \tilde{Q}^{tr}(\xi) \big) \notag
  \\
  &=\frac{1}{16\pi G_\text{N}} \oint_{r=\infty} \!\!\!\!\!\!\!\! d\phi \, 
  \Big\{ \xi^\rho \big( 2 e e^t{}_a e^r{}_b \om_\rho{}^{ab} 
  - \overline{\psi_\phi} \psi_\rho \big) + e \tilde{Q}^{tr}(\xi_1) \Big\}. \label{eq:Ham}
\end{alignat}
The equations of motion are imposed in the second line. 
Note that the Hamiltonian is written only by using the boundary values, 
so the behavior of the geometry (\ref{eq:AdS3}) is enough to discuss this quantity.
In order to check the Dirac bracket of the algebra, let us take the variation of the Hamiltonian.
\ba
  \dl_{\xi_2} H(\xi_1) = \{H(\xi_1), H(\xi_2) \} = H([\xi_1,\xi_2]) + K(\xi_1,\xi_2).
  \label{eq:braket1}
\end{alignat}
The last term is the central extension of the algebra. 

Now we evaluate this quantity in the background of the massless BTZ black hole,
eq.~(\ref{eq:AdS3}) with $\psi_\mu = 0$. Now the energy is adjusted so that $H(\xi) = 0$.
Thus $K(\xi_1,\xi_2) = \dl_{\xi_2} H(\xi_1)$ and from eq.~(\ref{eq:varSGcurrent}) we evaluate
\ba
  \dl_{\xi_2} H(\xi_1) 
  &=\frac{1}{16\pi G_\text{N}} \oint_{r=\infty} \!\!\!\!\!\!\!\! d\phi \, 
  \Big\{ \dl_{\xi_2} \big( e Q^{tr}(\xi_1) \big) 
  + e \big( \xi_1^t \Theta^r(\xi_2) - \xi_1^r \Theta^t(\xi_2) \big) \Big\} \notag
  \\
  &=\frac{1}{16\pi G_\text{N}} \oint_{r=\infty} \!\!\!\!\!\!\!\! d\phi \, 
  \Big\{ \xi_1^\rho \dl_{\xi_2} \big( 2 e e^t{}_a e^r{}_b \om_\rho{}^{ab} 
  - \overline{\psi_\phi} \psi_\rho \big) \notag
  \\
  &\qquad\qquad\qquad\quad\;\;
  + \xi_1^t \big( 2 e e^r{}_a e^\nu{}_b \dl_{\xi_2} \om_\nu{}^{ab} 
  + e \e^{r\nu\rho} \overline{\psi_\nu} \dl_{\xi_2} \psi_\rho \big) \notag
  \\
  &\qquad\qquad\qquad\quad\;\;
  - \xi_1^r \big( 2 e e^t{}_a e^\nu{}_b \dl_{\xi_2} \om_\nu{}^{ab} 
  + e \e^{t\nu\rho} \overline{\psi_\nu} \dl_{\xi_2} \psi_\rho \big) \Big\} \notag
  \\
  &=\frac{1}{16\pi G_\text{N}} \oint_{r=\infty} \!\!\!\!\!\!\!\! d\phi \, 2 e 
  \Big\{ e^\sg{}_c \dl_{\xi_2} e^c{}_\sg e^t{}_a e^r{}_b \xi_1^\rho \om_\rho{}^{ab} 
  + \dl_{\xi_2} e^t{}_a e^r{}_b \xi_1^\rho \om_\rho{}^{ab} \notag
  \\
  &\qquad\qquad\qquad\quad\quad\;\;
  + e^t{}_a \dl_{\xi_2} e^r{}_b \xi_1^\rho \om_\rho{}^{ab} 
  + e^t{}_a e^r{}_b \xi_1^\rho \dl_{\xi_2} \om_\rho{}^{ab}  \notag
  \\
  &\qquad\qquad\qquad\quad\quad\;\;
  + \xi_1^t e^r{}_a e^\nu{}_b \dl_{\xi_2} \om_\nu{}^{ab} 
  - \xi_1^r e^t{}_a e^\nu{}_b \dl_{\xi_2} \om_\nu{}^{ab} \Big\} \notag
  \\
  &= - \frac{\ell}{16\pi G_\text{N}} \oint_{r=\infty} \!\!\!\!\!\!\!\! d\phi \, 
  \big( 4 T_{1+} \pa_+^3 T_{2+} + 4 T_{1-} \pa_-^3 T_{2-} \big). \label{eq:Hcenter}
\end{alignat}
Here we used eq.~(\ref{eq:xiasym}) and the following relations.
\begin{alignat}{3}
  &e^\sg{}_c \dl_{\xi_2} e^c{}_\sg = 0, \qquad
  \dl_{\xi_2} e^t{}_a e^r{}_b \xi_1^\rho \om_\rho{}^{ab} = 0, \qquad
  \dl_{\xi_2} e^r{}_b = 0, \notag
  \\
  &e^t{}_a e^r{}_b \xi_1^\rho \dl_{\xi_2} \om_\rho{}^{ab} =
  - \frac{\ell}{r} \Big\{ 2 T_{1+} \pa_+^3 T_{2+} + 2 T_{1-} \pa_-^3 T_{2-} 
  + \big( \pa_+ T_{1+} + \pa_- T_{1-} \big) \big( \pa_+^2 T_{2+} + \pa_-^2 T_{2-} \big) \Big\}, \notag
  \\
  &e^\nu{}_b \dl_{\xi_2} \om_\nu{}^{ab} =
  \begin{pmatrix}
    \frac{1}{r} \big( \pa_+^2 T_{2+} + \pa_-^2 T_{2-} \big) \\
    - \frac{1}{r} \big( \pa_+^2 T_{2+} - \pa_-^2 T_{2-} \big) \\
    0
  \end{pmatrix},
  \\
  &\xi_1^t e^r{}_a e^\nu{}_b \dl_{\xi_2} \om_\nu{}^{ab} = 0, \notag
  \\
  &\xi_1^r e^t{}_a e^\nu{}_b \dl_{\xi_2} \om_\nu{}^{ab} =
  - \frac{\ell}{r} \big( \pa_+ T_{1+} + \pa_- T_{1-} \big) \big( \pa_+^2 T_{2+} + \pa_-^2 T_{2-} \big). \notag
\end{alignat}
Notice that left and right modes are separated in a nontrivial way in eq.~(\ref{eq:Hcenter}).
If we substitute Fourier mode expansions of eq.~(\ref{eq:xiasym2}), we obtain
\ba
  \dl_{\xi_{\pm,n}} H(\xi_{\pm,m}) &= - i \frac{\ell}{8 G_\text{N}} m^3 \dl_{m+n,0}.
\end{alignat}
This gives the central extensions of left and right Virasoro algebras.
By setting $H(\xi_{\pm,m}) = L^\pm_m e^{imx^\pm}$ and identifying $\{ \;,\; \}$ with $-i[ \;,\; ]$, 
the algebra (\ref{eq:braket1}) becomes
\begin{alignat}{3}
  [L^+_m, L^+_n] &= (m-n) L^+_{m+n} + \frac{c}{12} m^3 \dl_{m+n,0}, \notag
  \\
  [L^-_m, L^-_n] &= (m-n) L^-_{m+n} + \frac{c}{12} m^3 \dl_{m+n,0}, \label{eq:Virasoroalg}
\end{alignat}
where $c=\frac{3\ell}{2G_\text{N}}$ is the central charge.

Next we evaluate the Dirac bracket of the supercharge.
The charge for the super transformation is written as
\ba
  F(\e) &= \int d^2x \, e S^t(\e) \notag
  \\
  &= \frac{1}{16\pi G_\text{N}} \oint_{r=\infty} \!\!\!\!\!\!\!\! d\phi \, e U^{tr}(\e) \notag
  \\
  &= \frac{1}{16\pi G_\text{N}} \oint_{r=\infty} \!\!\!\!\!\!\!\! d\phi \, 2 \overline{\e} \psi_\phi. 
  \label{eq:supercharge}
\end{alignat}
The equations of motion are imposed in the second line. 
It is obvious that the supercharge is zero in the background of $\psi_\mu = 0$.
The variation of the supercurrent is evaluated as
\ba
  \delta_{\e_2} F(\e_1) &= \{ F(\e_1), F(\e_2) \} 
  = H(\overline{\e_1}\g \e_2) + K(\e_1,\e_2), \label{eq:Fcenter}
\end{alignat}
where $K(\e_1,\e_2)$ is the central extension of the algebra.
Now we evaluate this quantity in the background of the massless BTZ black hole.
Then $K(\e_1,\e_2) = \dl_{\e_2} F(\e_1)$ and from eq.~(\ref{eq:varSGsuper}) we obtain
\begin{alignat}{3}
  \delta_{\e_2} F(\e_1) 
  &= \frac{1}{16\pi G_\text{N}} \oint_{r=\infty} \!\!\!\!\!\!\!\! d\phi \,
  \big( 2 \overline{\e_1} \mathcal{D}_\phi \e_2 - 2 \overline{\e_2} \mathcal{D}_\phi \e_1 \big) \notag
  \\
  &= \frac{i \ell}{16\pi G_\text{N}} \oint_{r=\infty} \!\!\!\!\!\!\!\! d\phi \,
  \big( 2 \chi_1^T \chi''_2 - 2 \chi_2^T \chi''_1 \big). \label{eq:Fcenter2}
\end{alignat}
If we substitute Fourier mode expansions of eq.~(\ref{eq:easym2}), we obtain
\ba
  \delta_{\e_t} F(\e_s) &= -i \frac{\ell}{2G_\text{N}} s^2 \dl_{s+t,0}. 
\end{alignat}
This corresponds to the central extension of the super Virasoro algebra.
Notice that $\overline{\e_s} \g^\mu \e_t = - 2i \xi_{+,s+t}^\mu$. 
Then, by setting $F(\e_s) = G_s e^{isx^+}$, the algebra (\ref{eq:Fcenter}) is expressed as
\begin{alignat}{3}
  \{ G_s, G_t \} &= 2 L^+_{s+t} + \frac{c}{3} s^2 \dl_{s+t,0}.
\end{alignat}

Let us examine a consistency check of eq.~(\ref{eq:Fcenter}).
In order to check this expression, let us take the variation of $\delta_{\e_2} F(\e_1)$.
The calculation becomes as follows.
\ba
  \dl_{\xi_2} \delta_{\e_2} F(\e_1) 
  &= \frac{1}{16\pi G_\text{N}} \oint_{r=\infty} \!\!\!\!\!\!\!\! d\phi \,
  \dl_{\xi_2} \big( 2 \overline{\e_1} \mathcal{D}_\phi \e_2 - 2 \overline{\e_2} \mathcal{D}_\phi \e_1 \big) \notag
  \\
  &= \frac{1}{16\pi G_\text{N}} \oint_{r=\infty} \!\!\!\!\!\!\!\! d\phi \,
  \Big( \overline{\e_1} \g^{ab} \e_2 \dl_{\xi_2} \om_{\phi ab} 
  + \frac{2}{\ell} \overline{\e_1} \g_a \e_2 \dl_{\xi_2} e^a{}_\phi \Big) \notag
  \\
  &= \frac{1}{16\pi G_\text{N}} \oint_{r=\infty} \!\!\!\!\!\!\!\! d\phi \,
  \overline{\e_1} \g_a \e_2 \dl_{\xi_2} \Big( \e^{abc} \om_{\phi bc} + \frac{2}{\ell} e^a{}_\phi \Big) \notag
  \\
  &= - \frac{2i\ell}{16\pi G_\text{N}} \oint_{r=\infty} \!\!\!\!\!\!\!\! d\phi \,
  4 T_{1+} \pa_+^3 T_{2+} \notag
  \\
  &= \dl_{\xi_2} H(-2i \xi_{1+}),
\end{alignat}
where $\overline{\e_1} \g^\mu \e_2 = -2i \xi_{1+}^\mu$.
Eq.~(\ref{eq:Fcenter}) is correctly derived by integrating this equation.
The third line in the above equation indicates the connection between
three dimensional supergravity and the gauge Chern-Simons theory.

Finally let us investigate the Dirac bracket of the Hamiltonian and the supercharge.
By employing the results obtained in section \ref{sec:Killing}, it is possible to show the following equation.
\begin{alignat}{3}
  \dl_{\e_2} \dl_{\xi_+} F(\e_1) 
  &= \frac{1}{4\pi G_\text{N}} \oint_{r=\infty} \!\!\!\!\!\!\!\! d\phi \,
  \overline{\e_1} \Big\{ \pa_\phi \xi_+^\rho \mathcal{D}_\rho \e_2 + \xi_+^\rho \pa_\rho (\mathcal{D}_\phi \e_2)
  + \frac{1}{4} \Lambda_{ab} \g^{ab} \mathcal{D}_\phi \e_2 \Big\} \notag
  \\
  &= \frac{i\ell}{4\pi G_\text{N}} \oint_{r=\infty} \!\!\!\!\!\!\!\! d\phi \,
  \Big( 2 T_+ \chi_1^T \chi'''_2 + 3 T'_+ \chi_1^T \chi''_2 \Big) \notag
  \\
  &= - \dl_{\e_2} F(\dl_{\xi_+} \e_1), \label{eq:HF}
\end{alignat}
where $\xi_+$ is the general coordinate transformation of the asymptotic symmetry group 
which depends only on $T_+(x^+)$.
We also defined
\begin{alignat}{3}
  \dl_{\xi_+} \e_1 &= \xi_+^\rho \pa_\rho \e_1 + \frac{1}{4} \Lambda_{ab} \g^{ab} \e_1,
\end{alignat}
which satisfies $\dl_{\dl_{\xi_+} \e_1} = \dl_{\e_1} \dl_{\xi_+} - \dl_{\xi_+} \dl_{\e_1}$.
If we substitute eqs.~(\ref{eq:xiasym2}) and (\ref{eq:easym2}), the above equation becomes
\begin{alignat}{3}
  \dl_{\xi_{+,m}} \e_s &= -i\Big(\frac{m}{2} - s \Big) \e_{m+s}.
\end{alignat}
By integrating eq.~(\ref{eq:HF}), we obtain
\begin{alignat}{3}
  \dl_{\xi_+} F(\e_1) 
  &= \{ F(\e_1), H(\xi_+) \} 
  = - F(\dl_{\xi_+} \e_1).
\end{alignat}
Notice that the integral constant should be zero since $F(\e) = 0$ for $\psi_\mu = 0$.
By setting $\xi_+ = \xi_{+,m}$ and $\e_1 = \e_s$, eventually we get
\begin{alignat}{3}
  [ L_m^+, G_s ] = \Big(\frac{m}{2} - s \Big) G_{m+s}.
\end{alignat}
In a similar way it is possible to show $[ L_m^-, G_s ] = 0$.
Therefore we conclude that there exist a direct product of Virasoro algebras at the boundary,
and one of them is extended to super Virasoro algebra.



\section{Conclusion and Discussion}

In this paper, we have formulated the current for the general covariance and that for
the local supersymmetry in three dimensional $\mathcal{N}=(1,0)$ supergravity.
We employed Noether's method and constructed them in a covariant way 
in terms of the vielbein and the spin connection. In order to make the variations of
the currents consistent, we referred Wald's approach and defined the Hamiltonian and the supercharge.

We also examined the asymptotic supersymmetry group which preserves the boundary behavior
of AdS$_3$ geometry. 
We solved the relaxed Killing vector and Killing spinor equations and obtained explicit forms 
of the general coordinate transformation $\xi^\mu$ and the local supersymmetry $\e$.
$\xi^\mu$ consists of arbitrary functions of $T_{+}(x^+)$ and $T_{-}(x^-)$, and
$\e$ is written by an arbitrary function of $\chi(x^+)$.
The vector constructed from the bilinear of two Killing spinors correctly matches with 
the Killing vector, and we found NS and R sectors for the supersymmetric states.
As discussed in appendix \ref{app:Killing}, the global AdS$_3$ belongs to the NS sector
and massless BTZ black hole does to the R sector.

We evaluated the variation of the Hamiltonian under the general coordinate transformation.
This is written in terms of the Hamiltonian with central extension
by using $T_{+}(x^+)$ and $T_{-}(x^-)$.
We also calculated the variation of the supercharge under the local supersymmetry.
This is given in terms of the Hamiltonian with central extension by using $\chi(x^+)$. 
Inserting mode expansions of the Hamiltonian and the supercharge, we showed that there exists 
the direct product of super Virasoro algebra and Virasoro algebra at the boundary of AdS$_3$.

The Virasoro algebras of eq.~(\ref{eq:Virasoroalg}) are not canonical form, that is,
the central extensions are not like $\frac{c}{12}(m^3-m)$.
To make eq.~(\ref{eq:Virasoroalg}) canonical form, we just shift $L_0^\pm$ as
\begin{alignat}{3}
  L'{}_0^\pm = L_0^\pm + \frac{c}{24}. 
\end{alignat}
Then the energy of the global AdS$_3$ $\Delta_0$ is shifted from $-\frac{1}{8G_\text{N}}$ to 0.
And the effective central charge becomes $c_\text{eff} \equiv c - 24\Delta_0 = c$~\cite{carl2}.
Thus it is possible to estimate the entropy of the BTZ black hole by using Cardy formula.

The formulation developed in this paper is applicable to general supergravity theories~\cite{hms}.
For example, it is straight forward to generalize the procedure of this paper
to $\mathcal{N}=(1,1)$ supergravity.
As a future direction, it is interesting to apply our formulation to the chiral supergravity
which contains Lorentz Chern-Simons term~\cite{beck}. 
it is also interesting to develop an extension to higher spin supergravity~\cite{henn}.

\section*{Acknowledgement}

The author would like to thank Yuji Sugawara and Takahiro Nishinaka for useful discussions and comments.
This work is partially supported by the Ministry of Education, Science, 
Sports and Culture, Grant-in-Aid for Young Scientists (B), 12014331, 2012.

\appendix

\section{Notation of the Gamma Matrix}
\label{app:gamma}

The gamma matrix in three dimensions satisfy the Clifford algebra,
\ba
  \{\gamma^a, \gamma^b \} = 2 \eta^{ab},
\end{alignat}
where $\eta^{ab} = \text{diag}(-1,1,1)$ and $a, b = 0,1,2$ denote local Lorentz indices. 
A space-time dependent gamma matrix is given by $\g^\mu = e^\mu{}_a \g^a$,
and a completely antisymmetric tensor $\gamma^{\mu_1\cdots \mu_n}$ is defined 
so that a coefficient of each term becomes $1/n!$. 
For example, $\gamma^{\mu\nu} = \frac{1}{2!}(\gamma^\mu\gamma^\nu - \gamma^\nu \gamma^\mu)$.
In some calculations, we often need to evaluate products of antisymmetric tensors.
Here we present some of them.
\ba
  \g^{\mu\nu\rho} \g_{ab} &= - 6 \g^{[\rho\mu}{}_{[a} e^{\nu]}{}_{b]}
  - 6 \g^{[\rho} e^\mu{}_{[a} e^{\nu]}{}_{b]}, \notag
  \\
  \frac{1}{2} \{\g^{\mu\nu\rho}, \g_{ab}\} &= - 6 \g^{[\rho} e^\mu{}_{[a} e^{\nu]}{}_{b]}, \notag
  \\
  \g^{\mu\nu\rho} \g_{\mu\nu} &= - 2 \g^{\rho}, \label{eq:gm1}
  \\
  \g_\rho \g^{\mu\nu\rho} &= \g^{\mu\nu\rho} \g_\rho. \notag
\end{alignat}
Also we often use a relation for Majorana spinors $\chi$ and $\eta$,
\ba
  \overline{\chi} \g^{\mu_1 \cdots \mu_n} \eta = 
  (-1)^{\frac{n(n+1)}{2}} \overline{\eta} \g^{\mu_1 \cdots \mu_n} \chi.
\end{alignat}
In eq.~(\ref{eq:covD}), we introduced two kinds of covariant derivatives.
$D_\mu$ acts only on local Lorentz indices and satisfy $D_\mu e^a{}_\nu = \Gamma^a{}_{\mu\nu}$
and $D_\mu \g^a =0$. From this we see that $D_{[\mu} e^a{}_{\nu]} = \Gamma^a{}_{[\mu\nu]}$
and $D_\rho (e \g^{\rho\mu_1 \cdots \mu_n})$ becomes the order of $\mathcal{O}(\psi^2)$. 
Especially $D_\gamma (e \g^{\mu\nu\rho}) = 0$ in three dimensions.
It is helpful to note following relations.
\ba
  e \overline{D_\rho \chi} \g^{\mu\nu\rho} \eta &= 
  \partial_\rho \big( e \overline{\chi} \g^{\mu\nu\rho} \eta \big)
  - \overline{\chi} D_\rho \big( e \g^{\mu\nu\rho} \big) \eta 
  - e \overline{\chi} \g^{\mu\nu\rho} D_\rho \eta \notag
  \\
  &= \partial_\rho \big( e \overline{\chi} \g^{\mu\nu\rho} \eta \big)
  - e \overline{\chi} \g^{\mu\nu\rho} D_\rho \eta,
  \\
  e \overline{\mathcal{D}_\rho \chi} \g^{\mu\nu\rho} \eta &=
  \partial_\rho \big( e \overline{\chi} \g^{\mu\nu\rho} \eta \big)
  - e \overline{\chi} \g^{\mu\nu\rho} \mathcal{D}_\rho \eta. \notag
\end{alignat}
Commutation relations of each covariant derivative are evaluated as
\ba
  [D_\mu, D_\nu] &= \frac{1}{4} R_{ab\mu\nu} \g^{ab}, \notag
  \\
  [\mathcal{D}_\mu, \mathcal{D}_\nu] &= \frac{1}{4} R_{ab\mu\nu} \g^{ab} 
  + \frac{1}{2\ell^2} \g_{\mu\nu},
\end{alignat}
and covariant derivatives of the field strength of the Majorana gravitino
with all indices antisymmetrized are calculated like
\ba
  &D_{[\rho} \psi_{\mu\nu]} = \frac{1}{4} R_{ab[\mu\nu} \g^{ab} \psi_{\rho]}, \notag
  \\
  &\mathcal{D}_{[\rho} \psi_{\mu\nu]} = \frac{1}{4} R_{ab[\mu\nu} \g^{ab} \psi_{\rho]}
  + \frac{1}{2\ell^2} \g_{[\mu\nu} \psi_{\rho]}. 
\end{alignat}

\section{On the Spin Connection}
\label{app:spin}

Let us consider the variation of the Lagrangian (\ref{eq:SGLag}) and determine the spin connection
including torsion part. The calculation becomes as follows.
\ba
  &16 \pi G_\text{N} \dl \mathcal{L} \notag
  \\
  &= 2 e \Big( R^a{}_\mu - \frac{1}{2} e^a{}_\mu R - \frac{1}{\ell^2} e^a{}_\mu \Big) \dl e^\mu{}_a 
  - \frac{1}{2} e \overline{\dl \psi_\rho} \gamma^{\rho\mu\nu} \psi_{\mu\nu} 
  - e \overline{\psi_\rho} \gamma^{\rho\mu\nu} \mathcal{D}_\mu \dl \psi_\nu \notag
  \\
  &\quad\,
  - 2 e e^\mu{}_a e^\nu{}_b D_\nu \dl \om_\mu{}^{ab} 
  - \frac{1}{4} e \overline{\psi_\rho} \g^{\mu\nu\rho} \g_{ab} \psi_\nu \dl \om_\mu{}^{ab}
  - \frac{1}{2\ell} e \overline{\psi_\rho} \g^{\mu\nu\rho} \g_a \psi_\nu \dl e^a{}_\mu \notag
  \\[0.1cm]
  &= 2 e \tilde{G}^a{}_\mu \dl e^\mu{}_a + e \overline{\dl \psi_\rho} \Psi^\rho 
  + \pa_\mu \big( 2 e e^\mu{}_a e^\nu{}_b \dl \om_\nu{}^{ab} 
  + e \overline{\psi_\rho} \gamma^{\mu\rho\sigma} \dl \psi_\sigma \big) \notag
  \\
  &\quad\,
  + \Big( D_\nu \big( 2 e e^\mu{}_{[a} e^\nu{}_{b]} \big) 
  - \frac{1}{4} e \overline{\psi}_\rho \g^{\mu\nu\rho} \g_{ab} \psi_\nu \Big) \dl \om_\mu{}^{ab}.
  \label{eq:varsugra}
\end{alignat}
Equations of motion are defined as
\begin{alignat}{3}
  \tilde{G}^a{}_\mu &= R^a{}_\mu - \frac{1}{2} e^a{}_\mu R - \frac{1}{\ell^2} e^a{}_\mu 
  - \frac{1}{4\ell} e \overline{\psi_\rho} \g^{a\rho\sigma} \g_\mu \psi_\sigma, \notag
  \\
  \Psi^\rho &= - \gamma^{\rho\mu\nu} \psi_{\mu\nu} = - \e^{\rho\mu\nu} \psi_{\mu\nu}.
\end{alignat}
Since $\e_{\mu\nu\rho} \Psi^\rho = 2\psi_{\mu\nu}$, equation of motion for the Majorana gravitino is simply
written as $\psi_{\mu\nu}=0$. Each Term in front of $\dl \om_\mu{}^{ab}$ 
in eq.~(\ref{eq:varsugra}) are expressed as
\begin{alignat}{3}
  e^{-1} D_\nu \big( 2 e e^{\mu}{}_{[a} e^{\nu}{}_{b]} \big) &=
  2 e^\rho{}_c \partial_\nu e^c{}_\rho e^\mu{}_{[a} e^\nu{}_{b]}
  + 2 \partial_\nu e^{\mu}{}_{[a} e^{\nu}{}_{b]} + 2 \partial_\nu e^{\nu}{}_{[b} e^{\mu}{}_{a]} 
  - 2 \om_{[ab]}{}^{\mu} + 2 e^{\mu}{}_{[a} \om_{b]} \notag
  \\
  &= - 2 e^\mu{}_c \partial_\nu e^c{}_\rho e^{\rho}{}_{[a} e^\nu{}_{b]} 
  - 2 \om_{[ab]}{}^{\mu} 
  + 2 e^{\mu}{}_{[a} \hat{\om}_{b]},
  \\
  \frac{1}{4} \overline{\psi_\rho} \g^{\mu\nu\rho} \g_{ab} \psi_\nu &=
  - \frac{3}{2} \overline{\psi_\rho} \g^{[\rho} e^\mu{}_{[a} e^{\nu]}{}_{b]} \psi_\nu \notag
  \\
  &= - \frac{1}{2} \overline{\psi_a} \g^\rho \psi_\rho e^{\mu}{}_b 
  + \frac{1}{2} \overline{\psi_b} \g^\rho \psi_\rho e^\mu{}_a
  + \frac{1}{2} \overline{\psi_a} \g^\mu \psi_b, \label{eq:gm2}
\end{alignat}
where $\om_b \equiv \om_{\rho b}{}^\rho$ and 
$\hat{\om}_b \equiv \om_b + e^\rho{}_c \partial_\nu e^c{}_\rho e^\nu{}_{b}
- e^\nu{}_c \partial_\nu e^c{}_\rho e^{\rho}{}_{b}$.
Then the equation of motion for the spin connection becomes
\begin{alignat}{3}
  - 2 e^\mu{}_c \partial_\nu e^c{}_\rho e^{\rho}{}_{[a} e^\nu{}_{b]} 
  - 2 \om_{[ab]}{}^{\mu} + 2 e^{\mu}{}_{[a} \hat{\om}_{b]}
  = - \frac{1}{2} \overline{\psi_a} \g^\rho \psi_\rho e^{\mu}{}_b 
  + \frac{1}{2} \overline{\psi_b} \g^\rho \psi_\rho e^\mu{}_a
  + \frac{1}{2} \overline{\psi_a} \g^\mu \psi_b.
\end{alignat}
By multiplying $e^b{}_\mu$ we obtain $\hat{\om}_a = \frac{1}{2} \overline{\psi_a} \g^\rho \psi_\rho$,
and by substituting this relation to the above we obtain
\begin{alignat}{3}
  \om_{\alpha\beta\gamma} - \om_{\beta\alpha\gamma} = 2 e_{\gamma c} \partial_{[\alpha} e^c{}_{\beta]} 
  - \frac{1}{2} \overline{\psi_\alpha} \g_\gamma \psi_\beta, \label{eq:torsion}
\end{alignat}
where $\omega_{\gamma\alpha\beta} \equiv \omega_{\gamma ab} e^a{}_\alpha e^b{}_\beta$.
The spin connection is given by
\begin{alignat}{3}
  \om_{\g\alpha\beta} &= 
  - e_{\gamma c} \partial_{[\alpha} e^c{}_{\beta]} + e_{\alpha c} \partial_{[\beta} e^c{}_{\g]} 
  + e_{\beta c} \partial_{[\g} e^c{}_{\alpha]} 
  + \frac{1}{4} \overline{\psi_\alpha} \g_\gamma \psi_\beta + \frac{1}{2} \overline{\psi_{[\alpha}} \g_{\beta]} \psi_\g.
  \label{eq:spincon}
\end{alignat}

Notice that from eq.~(\ref{eq:torsion}), antisymmetric part of the affine connection is written as
\begin{alignat}{3}
  \Gamma^c{}_{[\alpha\beta]} &= D_{[\alpha} e^c{}_{\beta]}
  = \frac{1}{4} \overline{\psi_\alpha} \g^c \psi_\beta.
\end{alignat}
Taking the variation of this equation, we obtain
\begin{alignat}{3}
  D_{[\alpha} \delta e^c{}_{\beta]} + \frac{1}{2} e^d{}_{[\beta} \delta \omega_{\alpha]}{}^c{}_d 
  = \frac{1}{4} \overline{\delta \psi_{[\alpha}} \g^c \psi_{\beta]}.
\end{alignat}
The variation of the spin connection is obtained by a linear combination of these equations
with indices cyclically rotated.
\begin{alignat}{3}
  \dl \om_{\rho cd} e^c{}_\mu e^d{}_\nu &= 
  \dl \om_{\dot{\rho} cd} e^c{}_\mu e^d{}_{\dot{\nu}} 
  + \dl \om_{\dot{\mu} cd} e^c{}_\nu e^d{}_{\dot{\rho}} 
  - \dl \om_{\dot{\nu} cd} e^c{}_\rho e^d{}_{\dot{\mu}} \notag
  \\
  &= - e_{c\mu} D_{\dot{\rho}} \dl e^c{}_{\dot{\nu}} 
  - e_{c\nu} D_{\dot{\mu}} \dl e^c{}_{\dot{\rho}} 
  + e_{c\rho} D_{\dot{\nu}} \dl e^c{}_{\dot{\mu}} \notag
  \\
  &\quad\,
  + \frac{1}{2} \overline{\dl \psi_{\dot{\rho}}} \g_\mu \psi_{\dot{\nu}} 
  + \frac{1}{2} \overline{\dl \psi_{\dot{\mu}}} \g_\nu \psi_{\dot{\rho}} 
  - \frac{1}{2} \overline{\dl \psi_{\dot{\nu}}} \g_\rho \psi_{\dot{\mu}} \notag
  \\
  &= - e_{c\dot{\mu}} D_{\rho} \dl e^c{}_{\dot{\nu}} 
  + e_{c\dot{\mu}} D_{\dot{\nu}} \dl e^c{}_{\rho} 
  + e_{c\rho} D_{\dot{\nu}} \dl e^c{}_{\dot{\mu}} \notag
  \\
  &\quad\,
  + \frac{1}{2} \overline{\dl \psi_{\rho}} \g_{\dot{\mu}} \psi_{\dot{\nu}} 
  - \frac{1}{2} \overline{\dl \psi_{\dot{\nu}}} \g_{\dot{\mu}} \psi_{\rho}
  - \frac{1}{2} \overline{\dl \psi_{\dot{\nu}}} \g_{\rho} \psi_{\dot{\mu}}. \label{eq:varspin2}
\end{alignat}

\section{Killing Vector and Killing Spinor for AdS$_3$ Geometry}
\label{app:Killing}

Let us consider isometry of global AdS$_3$ geometry and BTZ black hole.
We will explicitly construct generators of the isometry for the global AdS$_3$ and
show that these form subalgebra of super Virasoro algebras.
We also construct generators of the isometry for the BTZ black hole and
show that these are time translation and angular rotation.
For the extremal case, it is possible to find a generator for the local supersymmetry.

The metric of BTZ black hole is given by
\begin{alignat}{3}
  ds^2 &= - N^2 dt^2 + r^2 (d\phi + N^\phi dt)^2 + N^{-2} dr^2,
  \\
  N^2 &= \frac{r^2}{\ell^2} + \Big(\frac{4G_\text{N}j}{r}\Big)^2 - 8 G_\text{N} m, \qquad
  N^\phi = \frac{4 G_\text{N}j}{r^2}. \notag
\end{alignat}
$m$ represents the mass of the black hole and $j$ does the angular momentum.
The black hole becomes extremal when $m = \frac{|j|}{\ell}$.
Global AdS$_3$ is realized when $m=-\frac{1}{8G_\text{N}}$ and $j=0$.

Let us consider the Killing vector equation $\dl_\xi g_{\mu\nu} = 0$
for the globally AdS$_3$ geometry. The solution of this equation is given by
\ba
  \xi &= a_1 e^{i x^+} \Big( \frac{r}{2N} \partial_t + \frac{\ell N}{2r} \partial_\phi
  - \frac{i\ell}{2} N \partial_r \Big) 
  + a_{-1} e^{-i x^+} \Big( \frac{r}{2N} \partial_t + \frac{\ell N}{2r} \partial_\phi
  + \frac{i\ell}{2} N \partial_r \Big) 
  + a_0 \partial_+ \notag
  \\
  &\quad
  + b_1 e^{i x^-} \Big( \frac{r}{2N} \partial_t - \frac{\ell N}{2r} \partial_\phi 
  - \frac{i\ell}{2} N \partial_r \Big) 
  + b_{-1} e^{-i x^-} \Big( \frac{r}{2N} \partial_t - \frac{\ell N}{2r} \partial_\phi 
  + \frac{i\ell}{2} N \partial_r \Big) 
  + b_0 \partial_- 
  \\
  &\equiv \sum_{n=-1,0,1} a_n L_n + \sum_{n=-1,0,1} b_n \tilde{L}_n, \notag
\end{alignat}
where $x^\pm = \frac{t}{\ell} \pm \phi$, 
$N=\sqrt{1+\frac{r^2}{\ell^2}}$ and $a_n$, $b_n$ are some constants which satisfy
$a_{-1}=a_1^\ast$, $b_{-1}=b_1^\ast$, $a_0^\ast = a_0$ and $b_0^\ast = b_0$.
$L_n$ and $\tilde{L}_n$ are properly normalized generators of the isometry 
and satisfy following commutation relations: 
\begin{alignat}{3}
  [L_m, L_n] = -i(m-n) L_{m+n}, \quad 
  [\tilde{L}_m, \tilde{L}_n] = -i(m-n) \tilde{L}_{m+n}.
\end{alignat}
Therefore the global AdS$_3$ possesses $SL(2,\mathbf{R}) \times SL(2,\mathbf{R})$ symmetry
of Virasoro subalgebras.

Since the BTZ black hole is constructed as a quotient space-time of global AdS$_3$,
it also possesses $SL(2,\mathbf{R}) \times SL(2,\mathbf{R})$ symmetry locally.
Actually it is possible to solve Killing vector equation and obtain 6 generators.
Among 6 generators, however, only time translation $\pa_t$ and rotation $\pa_\phi$
are consistent with the global structure of the geometry.
Thus locally AdS$_3$ does not possess symmetry of Virasoro subalgebras.

Next let us examine supersymmetry of the geometry. 
Here we choose the explicit form of the gamma matrix as
\ba
  \g^0 = \begin{pmatrix} 0 & -1 \\ 1 & 0 \end{pmatrix}, \quad
  \g^1 = \begin{pmatrix} 0 & 1 \\ 1 & 0 \end{pmatrix}, \quad
  \g^2 = \begin{pmatrix} -1 & 0 \\ 0 & 1 \end{pmatrix}. \quad
\end{alignat}
Note that $\g^{012} = \mathbf{1}$ and $\g^{ab} = \e^{abc} \g_c$.

Then Killing spinor equation $\dl_\e \psi_\mu = 0$ for the globally AdS$_3$ is expressed as
\begin{alignat}{3}
  &\delta_\epsilon \psi_t
  = 2 \partial_t \epsilon + \frac{1}{\ell} \Big( - N \g^0 + \frac{r}{\ell} 
  \g^1 \Big) \epsilon = 0 , \notag
  \\
  &\delta_\epsilon \psi_\phi
  = 2 \partial_\phi \epsilon + \Big( - N \g^0 + \frac{r}{\ell}
  \g^1 \Big) \epsilon = 0 , \label{eq:KS1}
  \\
  &\delta_\epsilon \psi_r
  = 2 \partial_r \epsilon + \frac{1}{\ell N} \g^2 \epsilon
  = 0. \notag
\end{alignat}
Here $\epsilon(x)$ is a space-time dependent Majorana spinor and solved as
\begin{alignat}{3}
  \epsilon &= c_{1/2} \frac{\ell}{2} e^{\frac{i}{2}x^+}
  \begin{pmatrix}
    A \\ - i A^{-1} 
  \end{pmatrix}
  + c_{-1/2} \frac{\ell}{2} e^{-\frac{i}{2} x^+}
  \begin{pmatrix}
    A \\ i A^{-1}
  \end{pmatrix} \label{eq:KS1sol}
  \\
  &\equiv c_{1/2} \e^+ + c_{-1/2} \e^{-}, \notag
\end{alignat}
where $c_{-1/2} = c_{1/2}^\ast$ and $A(r) = \sqrt{\frac{r}{\ell} + N}$.

Note that the Killing spinor becomes antiperiodic along the $\phi$ direction.
This means that the globally AdS$_3$ belongs to Neveu-Schwarz sector and 
preserves $\mathcal{N} = (1,0)$ local supersymmetry.
When $k_+ = 0$, BTZ black hole becomes extremal ($j=-m\ell$) and the Killing spinor is 
independent of the $\phi$ direction.
\begin{alignat}{3}
  \epsilon = \frac{\ell}{2}
  \begin{pmatrix}
    \sqrt{\frac{r}{\ell} - \frac{4G_\text{N}m\ell}{r}} \\
    0
  \end{pmatrix}.
\end{alignat}
This corresponds to Ramond sector.


\end{document}